\documentclass[prd,aps,twocolumn,nofootinbib,superscriptaddress,eqsecnum,floatfix,preprintnumbers,amsmath,amssymb,
nofootinbib,longbibliography]{revtex4-1}

\usepackage{amssymb,amsmath}
\usepackage{epsfig}
\usepackage[dvipsnames]{xcolor}
\usepackage[utf8]{inputenc}
\usepackage{stmaryrd}
\usepackage{mathrsfs}
\usepackage{mathalfa}
\usepackage{accents}
\usepackage[normalem]{ulem}
\usepackage{enumitem}

\usepackage{graphicx}

\newcommand{\JJ}{\mathcal{J}}
\newcommand{\PP}{\mathcal{P}}

\newcommand{\pa}{\partial}

\newcommand{\be}{\begin{equation}}
\newcommand{\ee}{\end{equation}}
\newcommand{\BS}{\begin{split}}
\newcommand{\ES}{\end{split}}
\newcommand{\bea}{\begin{eqnarray}}
\newcommand{\eea}{\end{eqnarray}}
\newcommand{\ba}{\begin{equation}\begin{aligned}}
\newcommand{\ea}{\end{aligned}\end{equation}}

\newcommand{\beg}{\begin{gather*}}
\newcommand{\eng}{\end{gather*}}
\newcommand{\hh}{,\hspace{0.5cm}}
\newcommand{\hhh}{,\hspace{0.2cm}}

\newcommand{\n}[1]{\label{#1}}

\newcommand{\CAL}{\mathcal}

\newcommand{\ts}[1]{{\boldsymbol{#1}}}

\usepackage[makeroom]{cancel}
\usepackage[caption=false]{subfig}
\usepackage[colorlinks=true,
            citecolor=green,
            linkcolor=red,
            filecolor=cyan,
            urlcolor=magenta,
            backref=false]{hyperref}

\newcommand{\ed}[1]{\ts{e}_{(#1)}}

\newcommand{\nnd}[1]{\ts{n}_{(#1)}}

\newcommand{\dual}[1]{{}^{\star\!}{#1}}

\begin{document}

\title{Motion of a rotating black hole in a homogeneous electromagnetic field}

\author{Valeri P. Frolov}%
\email[]{vfrolov@ualberta.ca}
\affiliation{Theoretical Physics Institute, Department of Physics,
University of Alberta,\\
Edmonton, Alberta, T6G 2E1, Canada
}

\begin{abstract}
In the present paper, we consider a rotating black hole moving in a static homogeneous electromagnetic field. We assume that the field is weak and neglect its backreaction on the geometry, so that the metric at far distance from the black hole is practically flat. We present an exact solution for a stationary electromagnetic field in the presence of the black hole for this problem and use it to calculate fluxes of the energy, momentum and angular momentum into the black hole. Using these results we derive the equations of motion of the rotating black hole in the electromagnetic field and discuss some of the interesting solutions of these equations. In particular, we demonstrate how the interaction of the spin of the black hole with the external magnetic field changes its trajectory.

\hfill {\scriptsize Alberta Thy 1-24}
\end{abstract}

\maketitle

\section{Introduction}

Study of the interaction of the electromagnetic field with black holes has long story. Using Newman-Penrose approach, Teukolsky \cite{teukolsky1972rotating,teukolsky1973perturbations} has demonstrated that the Maxwell equations in the Kerr metric allow complete separation of variables. This result gave powerful tools for study different effects connected with scattering and absorption of the electromagnetic waves by black holes. Well known examples of physically interesting effects are black hole shadow, superradiance and quasinormal modes. First pictures of supermassive black holes in the center of our Galaxy and in the galaxy M87 were obtained by the Event Horizon telescope by  observing the radiation emitted by plasma moving in their vicinity \cite{akiyama2019first,akiyama2021first,EHT:22}. The magnetic field surrounding black holes play an important role. In the Blandford–Znajek process \cite{blandford:77} the magnetic field extracts spin energy of a rotating black hole and transmits it to distant relativistic particles forming jets (see e.g. \cite{thorne1986black,thorne2017modern} and references therein).

In this paper we consider another aspect of the interaction of  black holes with the electromagnetic field. We consider a motion of a rotating black hole in a static homogeneous electromagnetic field and
discuss how such a field affects its motion and rotation. We assume that the field is weak and neglect its gravitational backreaction on the spacetime metric . This means that far from the black hole the spacetime is practically flat, while electric and magnetic fields are homogeneous in this domain.

If at least one of two invariants of the electromagnetic field does not vanish, then there exists a frame  in which the electric and magnetic  fields are parallel or one of these fields vanishes. We call it a rest frame of the field and denote it by $\tilde{K}$. We also denote by $\vec{B}_0$ and  $\vec{E}_0$ the parallel vectors of the electric and magnetic fields in this frame, respectively . We denote by $K$ a reference frame  moving with velocity $\vec{V}$ with respect to $\tilde{K}$ in which the black hole is at rest. Using the Lorentz transformation one can find the magnetic $\vec{B}$ and   electric $\vec{E}$ fields in the frame $K$. Using a solution of the Maxwell equations in the Kerr metric with asymptotics $\vec{B}$ and   $\vec{E}$, one can find energy-momentum and angular momentum fluxes of the electromagnetic field into the rotating black hole. This result allows one to obtain a force and torque acting on the black hole in $K$ frame.

Let us note that if the black hole does not rotate and is at rest in the field frame $\tilde{K}$  this problem is rather trivial. For the corresponding Schwarzschild black hole one can always choose its spherical coordinates $(r,\theta,\varphi)$ so that the axis of rotations generated by the Killing vector $\ts{\zeta}=\pa_{\varphi}$ at far distance coincides with  the direction of the parallel electric and magnetic fields. For this case the solution $F_{\mu\nu}$ for the magnetic field can be obtained from the 4D vector potential $A_{\mu}=B_0\zeta_{\mu}$.
Taking a dual field $\dual{F}_{\mu\nu}$ and changing $B_0$ to $E_0$ one gets a required solution for the asymptotically homogeneous electric field. It is easy to check that for the superposition of such electric and magnetic fields the fluxes of the energy, momentum and angular momentum into the non-rotating black hole vanish. This means that the black hole's spin always remains zero and its mass does not change. We shall show that this result remain valid when the non-rotating black hole moves with respect to $\tilde{K}$ frame. For this case the velocity  of its motion is constant as well.

As we show a situation when a black hole is rotating is much more interesting.
As a result of the motion of a rotating black hole in the homogeneous electromagnetic field, force arises acting on. There also exists  a torque acting on its spin\footnote{
One can expect an existence of the force and torque acting on a rotating black hole in the electromagnetic field in the framework of the membrane paradigm \cite{thorne1986black} in which the stretched horizon of the  black hole effectively behaves as a conducting surface. A rotation of a conducting sphere in the magnetic field induces eddy currents and their interaction with the magnetic field produces the torque. The interaction of these currents with the electric field generates  a  force acting of the sphere (see e.g. \cite{RS1,RS2,RS3} and references therein).
}. The goal of this paper is to calculate these quantities.

To calculate a force and torque acting on a moving rotating black hole we proceed as follows. First we derive these objects in the frame $\tilde{K}$ where the black hole is at rest. For a far distant observer such a black hole can be effectively described as a small-size spinning particle. A force and torque acting on the particle are related to the fluxes of the momentum and angular momentum of the electromagnetic field into the black hole, that change its momentum and spin.
To calculate these fluxes one needs to find a solution of the Maxwell equation in the background of a rotating black hole which is regular at the horizon and properly reproduces corresponding homogeneous at the infinity electric and magnetic fields. A stationary solution of the Maxwell equations is uniquely defined by these boundary conditions. A corresponding solution was obtained by Bičák \cite{Bicak:76} and discussed and studied in \cite{Bicak:85,Bicak:06,Karas:2000} (see also \cite{Bobo}). In our  derivation of the force acting on a rotating black hole moving in the static homogeneous electromagnetic field  we use a slightly modified version of this solution.

In order to describe a motion of the black hole and its spin evolution we use an adiabatic approximation.
Namely, we assume that a change of the black hole parameters due to the interaction with the electromagnetic field is slow and its metric can be effectively described by the Kerr metric with slowly changing parameters.
After  calculating the force and torque in the reference frame in which the black hole is instantly at rest,
we apply the Lorentz transformation and obtain these quantities in the rest frame of the field $\tilde{K}$.

In this paper we use the units in which $G=c=1$. In these units the stress-energy tensor of the electromagnetic field, which is a quadratic expression in terms of the field strength $F$, has the dimension $F^2\sim 1/length^2$.  The required fluxes of the energy-momentum are obtained by integrating the stress-energy tensor over a 2D surface surrounding the black hole. These fluxes are dimensionless  and  vanish when either the mass of the black hole  $M$ or its rotation parameter $a=J/M$ vanishes. This means that one can expect that such a flux will be of the form $\sim F^2 Ma$. Similarly,  the expected angular momentum flux should be a linear combination of two terms $\sim  F^2 Ma^2$ and $ F^2 M^2 a$. The results presented in this paper confirm these expectations\footnote{Presented expressions correctly reproduce the required dimension of the fluxes. Let us mention that in a general case formulas for the fluxes might contain functions of a dimensionless quantity $a/M$. The results of this paper show that in fact these functions reduce to  numbers.}.

Let us emphasize that in our calculations we use the exact solution for the Maxwell field in the Kerr metric and do not assume that the gravitational field of the black hole is weak or its rotation is slow. In this sense the problem we discussed in the paper is solved exactly and for this reason theoretically  it is quite interesting.
Let us note, that for a stellar mass black hole moving near a magnetized supermassive black hole the effects discussed in this paper are rather small. However, they might be important for the case of  small-mass primordial black holes moving in strong magnetic fields which can exist in the early Universe (see e.g. \cite{Grasso:2000wj}).

The paper is organized as follows. In section~II we collect formulas for the Kerr metric and expressions for the useful tetrad frames in it. In section~III we present a solution for the Maxwell equations in the Kerr metric which describes
a static electromagnetic field which is regular at the black hole horizon and homogeneous at the infinity. We  use this solution to calculate the energy-momentum and angular momentum fluxes of the electromagnetic field into the black hole.  In section~IV we derive the equations for black hole motion and its spin evolution and discuss some special solutions of these equations. Section~V contains discussion of the results obtained in the paper. A brief appendix contains expressions for Killing vectors in the flat spacetime in the oblate spheroidal coordinates, which are used in the paper.

\section{Kerr metric and ZAMO's frame}
\n{S2}
\subsection{The Kerr metric}

The Kerr metric, describing a vacuum stationary rotating black hole, written in the Boyer-Lindquist coordinates, is
\begin{equation}\label{Kerr}
\begin{split}
ds^2 &= -\left( 1-\frac{2Mr}{\Sigma}\right) dt^2
-\frac{4Mar\sin^2\theta}{\Sigma} dt d\varphi\\
&+\dfrac{A\sin^2\theta}{\Sigma} d\varphi^2+\frac{\Sigma}{\Delta} dr^2 +\Sigma d\theta^2\, ,\\
\Sigma&=r^2+a^2\cos^2\theta\hh \Delta=r^2-2Mr+a^2 \, ,\\
A&=(r^2+a^2)^2-a^2\Delta \sin^2\theta\\
&=\Sigma(r^2+a^2)+2M a^2 r\sin^2\theta\, .
\end{split}
\end{equation}
Here $M$ is the black hole mass, and $a$ is its rotation parameter related to the black-hole's spin $J=Ma$.
The metric has two commuting Killing vectors $\ts{\xi}=\pa_t$ and $\ts{\zeta}=\pa_{\phi}$. Let us denote
\be
r_{\pm}=M\pm \sqrt{M^2-a^2}\, .
\ee
Equation $r=r_+$, where $\Delta=0$, describes the event horizon. The surface area of the horizon is
\be
\CAL{A}=4\pi(r_+^2+a^2)=8\pi M r_+ \, .
\ee

The Boyer-Lindquist coordinates $(t,r,\theta,\varphi)$ are singular at the horizon. To describe both the exterior and interior of a rotating black hole one can use so called Kerr incoming coordinates $(v,r,\theta,\tilde{\varphi})$ which are regular at the future event horizon \cite{Kerr}
\ba
&dv=dt+dr_*\hh dr_*=(r^2+a^2)\dfrac{dr}{\Delta}\, ,\\
&d\tilde{\varphi}=d\varphi +a\dfrac{dr}{\Delta}\, .
\ea
Similarly, one can introduce Kerr outgoing coordinates $(u,r,\theta,\tilde{\varphi})$
\be
dv=dt-dr_*\hh d\tilde{\varphi}=d\varphi -a\dfrac{dr}{\Delta}\, ,
\ee
which are regular at the past horizon and cover the white hole domain.

For $M=0$ the curvature vanishes and the metric \eqref{Kerr} takes the form
\be\n{OBL}
ds^2=-dt^2+\dfrac{\Sigma}{r^2+a^2}dr^2+\Sigma d\theta^2 +(r^2+a^2)\sin^2\theta d\varphi^2\, .
\ee
This is nothing but a flat metric in the oblate spheroidal coordinates $(r,\theta,\varphi)$ related to the flat coordinates $(X,Y,Z)$ as follows
\be
\begin{split}
&X=\sqrt{r^2+a^2}\sin\theta\cos\varphi\, ,\\
&Y=\sqrt{r^2+a^2}\sin\theta\sin\varphi\, ,\\
&Z=r\cos\theta\, .
\end{split}
\ee
We denote by $\ts{\xi}_{(T)}$,  $\ts{\xi}_{(X)}$,  $\ts{\xi}_{(Y)}$, and  $\ts{\xi}_{(Z)}$ the Killing vectors generating translations along $t$, $X$, $Y$ and $Z$, and by $\ts{\zeta}_{(X)}$, $\ts{\zeta}_{(Y)}$ and $\ts{\zeta}_{(Y)}$ the Killing vectors generating rotations around $X$, $Y$ and $Z$ axes.
These vectors  in the oblate spheroidal coordinates are given in the Appendix.

\subsection{ZAMO's  frame}

 For the Kerr metric in the standard form \eqref{Kerr} there exists a preferable and useful choice of observers and tetrads: a so called frame of a Zero Angular Momentum Observer (ZAMO) \cite{bardeen1972rotating,thorne1986black}. The basis vectors of this tetrad are
 \be\n{ZAMO}
\begin{split}
&\ed{t}=\left(\dfrac{A}{\Sigma\Delta}\right)^{1/2}\dfrac{\pa}{\pa t} +\dfrac{2Mar}{(A\Sigma\Delta)^{1/2}}\dfrac{\pa}{\pa \varphi}\, ,\\
& \ed{r}=\left(  \dfrac{\Delta}{\Sigma}  \right)^{1/2}\dfrac{\pa}{\pa r}\, ,\\
& \ed{\theta}=\dfrac{1}{\Sigma^{1/2}}\ \ \dfrac{\pa}{\pa \theta}\, ,\\
& \ed{\varphi}=\left(\dfrac{\Sigma}{A}\right)^{1/2}\dfrac{\pa}{\pa \varphi}\, .
\end{split}
\ee

At each point one can make the following spatial rigid rotation of the basis vectors $(\ed{r},\ed{\theta},\ed{\varphi})$ and define 3  new unit and mutually orthogonal vectors
\be\n{nXYZ}
\begin{split}
&\nnd{X}=\sin \theta\cos\varphi\ed{r}+\cos\theta\cos\varphi \ed{\theta}-\sin\varphi \ed{\varphi}\, ,\\
& \nnd{Y}=\sin \theta\sin\varphi\ed{r}+\cos\theta\sin\varphi \ed{\theta}+\cos\varphi \ed{\varphi}\, ,\\
& \nnd{Z}=\cos\theta \ed{\theta}-\sin\theta \ed{\varphi}\, .
\end{split}
\ee
At large distance from the black hole these vectors has the following asymptotic form
\be
\nnd{X}=\dfrac{\pa}{\pa X}\hh \nnd{Y}=\dfrac{\pa}{\pa Y}\hh \nnd{Z}=\dfrac{\pa}{\pa Z}\, ,
\ee
where $(X,Y,Z)$ are flat (Cartesian) coordinates.

\section{A black hole in a homogeneous electromagnetic field}

\subsection{Electromagnetic field}

Let us consider a rotating black hole which is immersed in a homogeneous at the infinity magnetic field. Such a solution of the source-free Maxwell equations in the Kerr spacetime
\be \n{ME}
F^{\mu \nu}_{\ \ ;\nu}=0\hh F_{[\mu\nu ,\alpha]}=0
\, ,
\ee
was obtained in \cite{Bicak:76} and discussed and studied in \cite{Bicak:85,Bicak:06,Karas:2000}. We use the following expression for the 4D vector  potential $\ts{A}=(A_t,A_r,A_{\theta},A_{\varphi})$ of the electromagnetic field which is a slightly modified version of the expression presented in \cite{Bicak:85}
\be\n{AXYZ}
\begin{split}
A_t&=B_Z a (M r (1+\cos^2\theta)/\Sigma-1)\\
&+\dfrac{a M \sin\theta\cos\theta}{\Sigma}\big[
B_X  (r\cos\psi-a\sin\psi)\\
&+B_Y(r\sin\psi+a\cos\psi)\big]\, ,\\
A_r&=-(r-M)\cos\theta \sin\theta (B_X \sin\psi-B_Y \cos\psi )\, ,\\
A_{\theta}&=-(a r\sin^2\theta +a M\cos^2\theta )(B_X \cos\psi+B_Y \sin\psi )\\
&+[r^2\cos^2\theta -(M r-a^2)\cos(2 \theta)](B_X \sin\psi-B_Y \cos\psi )\, ,\\
A_{\varphi}&=B_Z  \sin\theta ^2 ((r^2+a^2)/2-a^2 M r (1+\cos^2\theta )/\Sigma) \\
&-\sin\theta \cos\theta\big[\Delta(B_X \cos\psi+B_Y\sin\psi)\\
&+\dfrac{M(r^2+a^2)}{\Sigma}\big(B_X (r\cos\psi -a\sin\psi )\\
&+B_Y(r\sin\psi +a\cos\psi )\big)\big]\, .
\end{split}
\ee
Here
\be \n{psi}
\psi=\phi+q(r)\hh q(r)=\dfrac{a}{\sqrt{M^2-a^2}}\ln\big( \dfrac{r-r_+}{r-r_-}
 \big)\, .
\ee

This potential depends on 3 parameters $B_X$, $B_Y$ and $B_Z$. We denote by $B_{\mu\nu}$ the strength of the field corresponding to this potential
\be
B_{\mu\nu}=A_{\nu ,\mu}-A_{\mu ,\nu}\, .
\ee
Let us denote
\be\n{FE}
E_{\mu\nu}=\dual{B}_{\mu\nu}\big|_{B_X\to E_X,B_Y\to E_Y,B_Y\to E_Y}\, ,
\ee
where
\be
\dual{B}_{\mu\nu}=\dfrac{1}{2} e_{\mu\nu\alpha\beta}B^{\alpha\beta}
\ee
is the tensor dual to the tensor $\ts{B}$.
A notation used in \eqref{FE} means that after the calculation of the tensor dual to $\ts{B}$, one needs to substitute new parameters $(E_X,E_Y,E_Z)$ instead of  $(B_X,B_Y,B_Z)$ . We denote by $\ts{F}$ a superposition of the fields $\ts{B}$ and $\ts{E}$
\be\n{FFFF}
F_{\mu\nu}=B_{\mu\nu}+E_{\mu\nu}\, .
\ee
One can check that a so-defined tensor $\ts{F}$ has the following properties
\begin{itemize}
\item $\ts{F}$ satisfies source-free Maxwell equations \eqref{ME};
\item It is regular at the black hole horizon;
\item Both electric $(Q)$ and magnetic monopole $(P)$ charges for this field vanish
\be \n{QP}
\begin{split}
& Q=\dfrac{1}{4\pi}\int_S\dual{F}^{\mu\nu} d\sigma_{\mu\nu}=0\, ,\\
&P=\dfrac{1}{4\pi}\int_S F^{\mu\nu} d\sigma_{\mu\nu}=0\, .
\end{split}
\ee
vanish.
\end{itemize}
In these relations $S$ is a  spacelike 2D surface surrounding the black hole.

Let us denote
\be\n{ELM}
\CAL{E}_{\mu}=\ed{t}^{\nu} F_{\nu\mu}\hh \CAL{B}_{\mu}=\ed{t}^{\nu}\dual{F}_{\nu\mu}\, ,
\ee
These 4D vectors $\ts{\CAL{E}}$ and  $\ts{\CAL{B}}$ has interpretation as the electric and magnetic fields as measured by ZAMO. The leading asymptotics of these vectors at large $r$ are of the form
\be
\begin{split}
&\ts{\CAL{E}}\approx E_X \nnd{X}+E_Y \nnd{Y}+E_Z \nnd{Z}\, ,\\
&\ts{\CAL{B}}\approx B_X \nnd{X}+B_Y \nnd{Y}+B_Z \nnd{Y}\, .
\end{split}
\ee
This means that the field $\ts{F}$ depends on 6 parameters $(E_X,E_Y,E_Z)$ and $(B_X,B_Y,B_Z)$, that are nothing, but the Cartesian components of constant and homogeneous at the infinity electric and  magnetic fields.

\subsection{Energy and angular momentum fluxes}

The stress-energy tensor of the electromagnetic field $\ts{F}$ is
\be\n{SET}
T_{\mu\nu}= \dfrac{1}{4\pi}  \big( F_{\mu\alpha} F_{\nu}^{\ \alpha}-\dfrac{1}{4}g_{\mu\nu} F_{\alpha\beta} F^{\alpha\beta} \big)\, .
\ee
Using two Killing vectors $\ts{\xi}$ and $\ts{\zeta}$ one can define two conserved currents
\be
\begin{split}\n{CONSERVE}
&\CAL{I}_{(\xi)}^{\mu}=-T^{\mu}_{\ \nu}\xi^{\nu}\hh
\CAL{I}_{(\zeta)}^{\mu}=T^{\mu}_{\ \nu}\zeta^{\nu}\, ,\\
&\CAL{I}_{(\xi)  ;\mu}^{\mu}=\CAL{I}_{(\zeta) ;\mu}^{\mu}=0\, .
\end{split}
\ee

Let us denote by $S_0$ a 2D surface defined by equations $t=t_0=$const , $r=r_0=$const and
denote by $\Sigma$ a 3D timelike surface describing the "evolution" of $S_0$ for the time interval $t_0\in (t_-,t_+)$.
The  volume element of the surface $\Sigma$ is
\be
d\sigma^{\mu}=-\delta_r^{\mu}\Delta \sin\theta\, dt\, d\theta \, d\varphi\, .
\ee

We denote the fluxes of the energy and angular momentum through 3D surface $\Sigma$  per a unit time $t$ by $\dot{E}$ and $\dot{J}$, respectively. Then one has
\be
\begin{split} \n{EJ_FLUX}
\dot{E}=&-\dfrac{1}{t_+-t_-}\int_{\Sigma} \xi^{\mu}T_{\mu\nu} d\sigma^{\mu}=\Delta \int_S T_{tr} d\omega\, ,\\
\dot{J}=&\dfrac{1}{t_+-t_-}\int_{\Sigma} \zeta^{\mu}T_{\mu\nu} d\sigma^{\mu}=-\Delta \int_S T_{\varphi r} d\omega\, ,\\
&d\omega=\sin\theta d\theta\, d\varphi \, ,\\
\end{split}
\ee
 The signs in these expressions are chosen so that these quantities describe the flux into the surface $S_0$ from its exterior. Using properties \eqref{CONSERVE} and the Stockes' theorem it is possible to show that the quantities $\dot{E}$ and $\dot{J}$ do not depend on the radius $r_0$ of the surface $S_0$ (see appendix~B of \cite{Frolov:2023gsl}).

Since $g_{tr}$ and $g_{r\varphi}$ components of the Kerr metric in the Boyer-Lindquist coordinates vanish one has
\be
T_{tr}=\dfrac{1}{4\pi}F_{t \alpha}F_{r}^{\ \alpha}
\hh T_{\varphi r}=\dfrac{1}{4\pi}F_{\varphi \alpha}F_{r}^{\ \alpha}  \, .
\ee
 Calculating integrals in \eqref{EJ_FLUX} one obtains
\be
\n{mj}
\dot{E}=0\hh \dot{J}=-\dfrac{2}{3} M^2 a (E_X^2+E_Y^2+B_X^2+B_Y^2)\, .
\ee
The first of this relation shows that the mass of the black hole does not change, while the second relation implies that the spin of the black hole cannot increase.

\subsection{Momentum and angular momentum fluxes}

Let $\ts{\eta}$ be a vector satisfying the relations $\CAL{L}_{\xi}\ts{\eta}=\CAL{L}_{\zeta}\ts{\eta}=0$. We use it  define the following expression
\be
 \n{XYZ_PP}
\CAL{P}[\ts{\eta}]=\dfrac{1}{t_+-t_-}\int_{\Sigma} \eta^{\mu}T_{\mu\nu} d\sigma^{\nu} .
\ee
For the Killing vectors $\ts{\xi}$ and $\ts{\zeta}$ this object coincides with expressions \eqref{EJ_FLUX} for the fluxes of the energy and  angular momentum. These quantities do not depend on the choice of the 2D surface $S_0$ surrounding the black hole.

To define the momentum and angular momentum fluxes of the electromagnetic field into the black hole
we use expression \eqref{XYZ_PP} in which we identify $\ts{\eta}$ with vectors  $\xi_{(a)}$ and $\zeta_{(a)}$ given in \eqref{TRANS} and \eqref{ROT}.
In the absence of a black hole, that is when $M= 0$, the vectors $\xi_{(a)}$ and $\zeta_{(a)}$ are exact Killing vectors. However, when $M\ne 0$ they are close to the Killing vectors only far from the black hole in the asymptotic domain. For this reason,  we assume that $r_0$ is large and take the limit $r_0\to\infty$. Using this prescription we write
\be\n{XYZ_infty}
\begin{split}
&\PP_{(a)}=-\lim_{r_0\to\infty}\left[\Delta \int_S T_{\mu \nu} \xi_{(a)}^{\mu} \delta^{\nu}_{r} d\omega\right] \, ,\\
&\JJ_{(a)}=-\lim_{r_0\to\infty}\left[\Delta \int_S T_{\mu \nu} \zeta_{(a)}^{\mu} \delta^{\nu}_{r} d\omega\right] \, .
\end{split}
\ee
We interpret these quantities as the fluxes of the momentum and angular momentum of the electromagnetic field into the black hole\footnote{The spacetime of the Kerr black hole is asymptotically flat and hence besides the exact symmetries, generated by its Killing vectors, it also possesses so called asymptotic symmetries \cite{sachs1962asymptotic}. One can use them to define asymptotically conserved currents (see  e.g. in \cite{Dray:85,Bonga:20} and references therein). However, in the application of this approach to the electromagnetic field it is usually assumed that this field decreases rapidly enough at the infinity (see \cite{Bonga:20}). This condition is violated for the homogeneous at the infinity field which we consider in this paper.}.

To calculate the quantities $\PP_{(a)}$ and $\JJ_{(a)}$ we use the GRTensor program. Let us note, that for these calculations it is possible to use $1/r$ expansion  of the stress-energy tensor. For  $\PP_{(a)}$ it is sufficient to keep terms up to the order $1/r^2$ of this expansion, while for $\JJ_{(a)}$ one should keep terms up to $1/r^3$ order. This means that  in the calculations it is also sufficient to keep terms of the electromagnetic field up to the third order in $1/r$ expansion. Let us mention that the components of the electromagnetic field depend on the angle variable $\psi=\phi+q(r)$, \eqref{psi}. After the components of the field $F_{\mu\nu}$  are calculated one should substitute this expression for $\psi$ in terms of the angle $\varphi$ and the function $q(r)$ given by \eqref{psi}.
In fact, it is sufficient to use only the first few terms of the expansion of the $cos(q(r))$ and $\sin(q(r))$ at large $r$.

Following this simplification rules and performing rather long calculations one obtains the following results.
The momentum fluxes are
\be
\begin{split} \n{XYZ_FLUX}
\PP_{(X)}=&-\dfrac{2}{3}Ma(B_X E_Z-B_Z E_X)\, ,\\
\PP_{(Y)}=&-\dfrac{2}{3}Ma(B_Y E_Z-B_Z E_Y)\, ,\\
\PP_{(Z)}=&0\, .
\end{split}
\ee

The fluxes of the angular momentum are
\be
\begin{split}\n{ANG}
\JJ_{(X)}=&-\dfrac{2}{3}M^2 a (B_X B_Z+E_X E_Z)\\
&-\dfrac{22}{15}Ma^2 (B_Y B_Z+E_Y E_Z)\, ,\\
\JJ_{(Y)}=&-\dfrac{2}{3}M^2 a (B_Y B_Z+E_Y E_Z)\\
&+\dfrac{22}{15}Ma^2 (B_X B_Z+E_X E_Z)\, ,\\
\JJ_{(Z)}=& -\dfrac{2}{3}M^2 a (E_X^2+ E_Y^2+B_X^2+ B_Y^2)\, .
\end{split}
\ee
The last expression correctly reproduces \eqref{mj}, as it should be.
The reason is that the asymptotic Killing vector $\zeta_Z^{\mu}$ generating a rotation around $Z$ axis in the flat metric  \eqref{OBL} written in the oblate coordinates coincides with the exact Killing vector $\ts{\zeta}$ of the Kerr metric in the Boyer-Lindquist coordinates.

Let us emphasize, that since the vectors $\ts{\xi}_{(a)}$ and $\ts{\zeta}_{(a)}$ in the limit $M\to 0$ coincide with the Killing vectors, the quantities $\ts{\PP}_{(a)}$ and $\ts{\JJ}_{(a)}$ vanishes as well in this limit.
Both fluxes vanish also for $a=0$. This means that a non-rotating (Schwarzschild) black hole  absorbs neither energy-momentum nor angular momentum, and its mass, velocity and spin are constant.
Let us emphasize, that for the rotating black hole the momentum flux is proportional to the factor $J=Ma$, while the fluxes of the angular momentum are linear combinations of two quantities proportional to $M^2a$ and $Ma^2$. For slowly rotating black holes when $a\ll M$ the second contribution is smaller than the first one.

\section{Equations of motion of a rotating black hole}

\subsection{Vector form of the equations of motion}

Consider an observer located at far distance $L\gg M$ from the black hole. For such an observer the black hole is a "small size" particle and  to describe its motion  one can use a "point-like" approximation.
Energy, momentum and angular momentum of the black hole change as a result of the fluxes. One can use relations \eqref{XYZ_FLUX} and \eqref{ANG} to define the 4D force $\ts{\CAL{F}}=( \CAL{F}^{t},\vec{\CAL{F}})$ and 3D torque $\vec{\CAL{T}}$ acting on the black hole.
These expressions are written in the rest frame of the black hole where the asymptotic Cartesian coordinates are chosen so that $Z-$axis coincides with the direction of the black hole's spin. It is convenient to write them in the coordinate independent form using vector notations
\be\n{VEC}
\begin{split}
\CAL{F}^t&=0\, ,\vec{\CAL{T}}=\vec{\CAL{T}}_1+\vec{\CAL{T}}_2\, ,\\
\vec{\CAL{F}}&=-\dfrac{2}{3}\vec{J}\times(\vec{B}\times\vec{E})\\
&=-\dfrac{2}{3}\Big( (\vec{J}\cdot\vec{E})\ \vec{B}- (\vec{J}\cdot\vec{B})\ \vec{E}\Big)\, ,\\
\vec{\CAL{T}}_1&=-\dfrac{2M}{3}\Big( \vec{B}\times (\vec{J}\times\vec{B})+\vec{E}\times (\vec{J}\times\vec{E})\Big)\, ,\\
\vec{\CAL{T}}_2&=\dfrac{22}{15 M}\Big(  (\vec{J}\cdot \vec{B}) ( \vec{J}\times\vec{B})+  (\vec{J}\cdot \vec{E})  ( \vec{J}\times\vec{E})\Big)\, .
\end{split}
\ee

\subsection{Rotating black hole motion in a constant electromagnetic field}

Expressions in \eqref{VEC} are written in the frame $K$ in which the black hole is at rest.
However, if the force acting on it does not vanish, the black hole begins to move. To describe this motion one can use the rest frame of the field $\tilde{K}$. If the black hole at a given moment of time moves with the velocity $\vec{V}$  with respect to $\tilde{K}$ then one can use a Lorentz transformation relating the two inertial frames: the frame $K$ instantly co-moving with the black hole and $\tilde{K}$.
For the components of the 4D force $\ts{f}=(f^0,\vec{f})$ in $\tilde{K}$ frame one gets
\be\n{fff}
\begin{split}
&f^0=\gamma (\vec{V}\cdot \vec{\CAL{F}}) \, ,\\
&\vec{f} =\vec{\CAL{F}}+(\gamma-1)\dfrac{ \vec{V}\cdot\vec{\CAL{F}}}{V^2}\vec{V}\, .\,
\end{split}
\ee
Here $\gamma=(1-V^2)^{-1/2}$.

In  $\tilde{K}$ frame
\be
\vec{B}_0=B_0\vec{n}\hh \vec{E}_0=E_0\vec{n}\, ,
\ee
where $B_0$ and $E_0$ are asymptotic values of the magnetic and electric fields, and $\vec{n}$ is their common direction. In the instantly co-moving frame $K$ the vectors $\vec{E}$ and $\vec{B}$ are
\be\n{EEBB}
\begin{split}
&\vec{E}=\gamma (\vec{E}_0+\vec{V}\times\vec{B}_0) -(\gamma-1)\dfrac{\vec{E}_0\cdot \vec{V}}{V^2}\vec{V}\, ,\\
&\vec{B}=\gamma (\vec{B}_0+\vec{V}\times\vec{E}_0) -(\gamma-1)\dfrac{\vec{B}_0\cdot \vec{V}}{V^2}\vec{V}\, .\,
\end{split}
\ee

During the motion the mass $M$ of the black hole remains constant.
The equations of motion of the black hole in $\tilde{K}$ frame are of the form
\be\n{BHMOV}
M\dfrac{d\gamma}{d\tau}=f^0\hh M \dfrac{d(\gamma\vec{V})}{d\tau}=\vec{f}\ .
\ee
Here $\tau$ is the proper time parameter along the black-hole's worldline. Since in the instantly co-moving frame $\tau$ coincides with the coordinate time $t$, one can write the equation for the spin evolution in the form
\be \n{SPIN}
\dfrac{d\vec{J}}{d\tau}=\vec{\CAL{T}}\, .
\ee
The set of equations \eqref{BHMOV} and \eqref{SPIN} describes the black hole moving in the constant homogeneous at the infinity electromagnetic field.

\subsection{Evolution of the black hole parameters}

As we mentioned relation $\CAL{F}^t=0$ in \eqref{VEC} implies that the mass of the black hole does not change. Let us show that value of its spin   either increases or remains constant. For this purpose we use the following relation
\be \n{DDJJ}
\dfrac{d J^2}{d\tau}=2 \vec{J}\cdot \dfrac{d\vec{J}}{d\tau}\, .
\ee
Using expression \eqref{VEC} for the torque component $\vec{\CAL{T}}_2$ one can conclude that $\vec{J}\cdot\vec{\CAL{T}}_2=0$. Using expression for $\vec{\CAL{T}}_1$ one gets
\be \n{JJJ}
\vec{J}\cdot\vec{\CAL{T}}_1=J^2 B^2-(\vec{J}\cdot\vec{B})^2+J^2 E^2-(\vec{J}\cdot\vec{E})^2\, .
\ee
Let $\vec{n}$, $\vec{b}$ and $\vec{e}$ be unit vectors in the direction of $\vec{J}$, $\vec{B}$ and $\vec{E}$, respectively
\be 
\vec{J}=J\vec{n}\hhh \vec{B}=B\vec{b}\hhh \vec{E}=E\vec{e}\, .
\ee
Then using \eqref{DDJJ} and \eqref{JJJ} one gets
\be 
\dfrac{d J^2}{d\tau}=2 J^2\Bigg[ B^2\Big( 1-(\vec{n}\cdot\vec{b})^2\Big)+E^2\Big( 1-(\vec{n}\cdot\vec{e})^2\Big)
\Bigg]\, .
\ee
The right-hand side of this relation is non-negative. If $EB\ne 0$, then it vanishes if the vectors  $\vec{n}$, $\vec{b}$ and $\vec{e}$ are parallel. Thus the spin of the black hole moving in the homogeneous electromagnetic field cannot increase. Since the mass $M$ is constant, the latter property implies that the radius $r_+$ of the black hole horizon
\be 
r_+=M+\sqrt{M^2-J^2/M^2}
\ee
never decreases as well. This means that the black hole area
\be 
\CAL{A}=8\pi Mr_+\, ,
\ee
and hence the black hole entropy, do not decrease, as it should be.

In the next subsection we obtain solutions of equations of motion of the black hole in the electromagnetic field for some interesting cases.

\subsection{Special cases}

\subsubsection{Rotating black hole at rest in $\widetilde{K}$ frame}

Let us consider first a simplest case when the black hole is at rest with respect to  $\tilde{K}$ frame, so that $\vec{B}=B_0\vec{n}$ and $\vec{E}=E_0\vec{n}$. We denote by $\vec{J}_{\parallel}$ and $\vec{J}_{\perp}$ parts of the spin vector which are parallel and orthogonal to $\vec{n}$, respectively
\be
\vec{J}_{\parallel}=J_n\vec{n}\hh J_n=(\vec{n}\cdot \vec{J})\hh \vec{J}_{\perp}=\vec{J}-J_n\vec{n}\, .
\ee
Using \eqref{SPIN} it is easy to check that
\be
\dfrac{dJ_n}{d\tau}=0\,\hh \dfrac{d\vec{J}_{\perp}}{d\tau}=-\dfrac{2M}{3}(B_0^2+E_0^2)\vec{J}_{\perp}\, .
\ee
The first of these relations shows that the spin projection on the direction of the fields is constant, while the transverse component $\vec{J}_{\perp}$ exponentially decreases with time
\be
\vec{J}_{\perp}=\vec{J}_{\perp}^0 \exp\big[-\dfrac{2M}{3}(B_0^2+E_0^2)\tau\big]\, .
\ee

\subsubsection{Motion of a rotating black hole transverse to the magnetic field }

As another example let us consider a motion of a rotating black hole in a homogeneous magnetic field. For this case the electric field in $\tilde{K}$ frame vanishes,  $\vec{E}_0=0$, and $\vec{B}_0=B_0 \vec{n}$. We also assume that
\be \n{VVJJ}
\vec{V}\cdot\vec{n}=0\hh \vec{J}=J\vec{n}\, .
\ee
Under these assumptions the following relations are valid
\be\n{OOO}
\begin{split}
&\vec{B}=\gamma B_0 \vec{n}\hhh \vec{E}=\gamma B_0 \vec{V}\times\vec{n}\hhh
\vec{V}\times\vec{\cal{F}}=0
\, ,\\
&\vec{E}\cdot\vec{n}=\vec{J}\times\vec{B}=0\hhh
\vec{J}\cdot\vec{B}=\gamma B_0 J\, ,\\
&\vec{J}\times\vec{E}=\gamma B_0 J \vec{V}\,  .
\end{split}
\ee
Using these relations one obtains the following expressions for $\vec{\CAL{F}}$ and $\vec{\CAL{T}}$
\be\n{FFTT}
\begin{split}
&\vec{\CAL{F}}=\dfrac{2}{3}\gamma^2 B_0^2 J\,  \vec{V}\times\vec{n}\, ,\\
&\vec{\CAL{T}}=\vec{\CAL{T}}_1=-\dfrac{2M}{3}\gamma^2 B_0^2 J V^2 \vec{n}\, ,
\end{split}
\ee
and \eqref{fff} gives
\be\n{nnn}
f^0=0\hh \vec{f}=\vec{\CAL{F}}=\dfrac{2}{3}\gamma^2 B_0^2 J (\vec{V}\times \vec{n})\, .
\ee

These equations imply that if the conditions \eqref{VVJJ} are valid at some initial moment of time, they are  valid for any later time as well.
Such a black holes moves in the plane orthogonal to the direction of the magnetic field and its spin is directed along the magnetic field.

Since $f^0=0$ the first of the equations \eqref{BHMOV} implies that the value $V$ of the velocity of the black hole is constant.
Using this property on can easily integrate the equation for the spin \eqref{SPIN}
with the following result
\be
\begin{split}
&\vec{J}=J(\tau) \vec{n}\hh J(\tau)=J_0\exp(-\Gamma \tau)\, ,\\
&\Gamma=\dfrac{2}{3}\gamma^2 V^2 B_0^2 M\,  .
\end{split}
\ee
This result means that for a moving rotating black hole  the spin along the magnetic field decreases. The characteristic time of this process is $\sim \Gamma^{-1}$.

The equation \eqref{BHMOV} for  the transverse to the field velocity takes the form
\be\n{VVJJBB}
\dfrac{d\vec{V}}{d\tau}=\omega (\vec{V}\times \vec{n})\hh
\omega=\dfrac{2\gamma B_0^2 J}{3M}\, .
\ee
To solve this equation we denote
\be
V_x+i V_y=V e^{-i\alpha(\tau)},
\ee
then the equation \eqref{VVJJBB} implies
\be
\dfrac{d\alpha}{d\tau}=\omega\, .
\ee
Integration of this equation gives
\be \n{alpha}
\alpha=\dfrac{J_0}{\gamma M^2V^2}\big[ 1-\exp(-\Gamma\tau)\big]\, .
\ee

For  a slow change  of the spin, that is when $\Gamma\ll 1/M$, one has
\be
\alpha\approx \omega \tau\, ,
\ee
where $\omega$ is given \eqref{VVJJBB}.
In this approximation the black hole revolves in the plane orthogonal to the magnetic field
along a circle of a slowly increasing radius
\be
R\approx\dfrac{V}{\omega}=\dfrac{3MV}{2\gamma B_0^2 J}\, .
\ee

\section{Discussion}

In this paper we discuss interaction of a rotating black hole with a static electromagnetic field. We assume that this field is homogeneous at far distance from the black hole. In this domain one can always find a reference frame in which either the electric and magnetic fields are parallel  or one of these fields vanishes. We denote such a frame by $\tilde{K}$ and discuss effects which arise when a black hole moves with respect to this frame. We showed that for a non-rotating black hole its parameters do not change and its velocity remains constant.

The case of a rotating black hole is more interesting. The mass of moving rotating black hole still remains constant, however its interaction with the electromagnetic field leads to the appearance of a force acting on the black hole and a torque which changes the value and orientation of its spin. Using the exact solution of the Maxwell equations in the Kerr spacetime, we calculated both, the force and torque, and derived the equation of motion which determines dynamics of the rotating black hole in the homogeneous electromagnetic field.

We present solutions of these equations for two cases. In the first case we assume that the black hole is at rest in the electromagnetic field in $\tilde{K}$ frame,  while its spin is tilted with respect to the common direction $\vec{n}$ of the electric and magnetic fields. We showed that the projection of the spin vector on $\vec{n}$ is constant, while the transverse to this direction component of the spin exponentially decreases. This effect is  known and described in the literature (see e.g. \cite{thorne1986black}).

In the second case we analysed a motion of the black hole in the magnetic field. Under the assumption that the velocity of the black hole is orthogonal to the field and its spin is parallel to the field,
we have integrated the equations of motion. We demonstrated that the value of the velocity $V$ of the black hole is constant, while its spin exponentially decreases with time. We calculated the corresponding characteristic time for the spin decay. By solving the equations of motion we demonstrated  that such a
black hole moves in the plane orthogonal to the magnetic field along circles with slowly decreasing angular frequency. The radius of the black hole orbit slowly increases.

It is interesting to compare the obtained results with the results of the recent paper \cite{Frolov:2023gsl} in which the motion of the rotating black hole in the homogeneous scalar field was considered. In the latter paper a special solution of the massless scalar field equation was used which has a constant non-vanishing timelike  gradient. The stress-energy tensor for such a fields in the flat background is constant. When the black hole moves with respect to this frame the force and torque also arise. However, there exists a big difference between the cases of the  scalar and electromagnetic field. As we showed,  the mass of the black hole in the electromagnetic field does not change, while for the scalar field it grows and, at least formally, it becomes infinite in the finite interval of time. It is interesting to understand such a big difference between these two cases. Naively, one can expect that this difference is connected with the fact that the vector of the scalar field's strength for the field discussed in \cite{Frolov:2023gsl} is a timelike vector, while for the electromagnetic field both vectors characterizing the field's strength, $\vec{E}$ and $\vec{B}$, are spacelike. In the recent paper \cite{FRKO} it is shown that when the gradient of the homogeneous scalar field is spacelike, the mass of a moving rotating black hole remains finite.

\appendix

\section{Killing vectors in the oblate spheroidal coordinates}

The Killing vectors generating time translations and translations along $X$, $Y$ and $Z$ axes
written in the oblate spheroidal coordinates $(r,\theta,\phi)$
 are
\be\n{TRANS}
\begin{split}
\xi_{(t)}^{\mu}\pa_{\mu}&=\pa_t\, ,\\
\xi_{(X)}^{\mu}\pa_{\mu}&=\dfrac{\sqrt{r^2+a^2}}{\Sigma}\cos\varphi \Big( r\sin\theta\ \pa_r+\cos\theta\ \pa_{\theta} \big) \\
&-\dfrac{\sin\varphi}{\sqrt{r^2+a^2}\sin\theta}\ \pa_{\phi}\, ,\\
\xi_{(Y)}^{\mu}\ \pa_{\mu}&=\dfrac{\sqrt{r^2+a^2}}{\Sigma}\sin\varphi \Big(r\sin\theta\ \pa_r+\cos\theta\ \pa_{\theta} \big) \\
&+\dfrac{\cos\varphi}{\sqrt{r^2+a^2}\sin\theta}\ \pa_{\phi}\, ,\\
\xi_{(Z)}^{\mu}\pa_{\mu}&=\dfrac{(r^2+a^2)\cos\theta}{\Sigma}\ \pa_r -\dfrac{r\sin\theta}{\Sigma}\ \pa_{\theta}\, .\\
\end{split}
\ee

The Killing vectors generating rotations around  $X$, $Y$ and $Z$ axes written in the oblate spheroidal coordinates $(r,\theta,\phi)$ are
\be\n{ROT}
\begin{split}
\zeta_{(X)}^{\mu}\pa_{\mu}&=\dfrac{\sqrt{r^2+a^2}}{\Sigma}\sin\varphi
\big[ a^2\sin\theta\cos\theta\ \pa_r-r\pa_{\theta}\big]\\
 & -\dfrac{r\cos\theta\cos\varphi}{\sqrt{r^2+a^2}\sin\theta}\ \pa_{\varphi}\, ,\\
\zeta_{(Y)}^{\mu}\pa_{\mu}&=-\dfrac{\sqrt{r^2+a^2}}{\Sigma}\cos\varphi
\big[a^2 \sin\theta\cos\theta\ \pa_r- r\pa_{\theta} \big] \\
& -\dfrac{r\cos\theta \sin\varphi }{\sqrt{r^2+a^2}\sin\theta}\ \pa_{\varphi}\, ,\\
\zeta_{(Z)}^{\mu}\pa_{\mu}&=\pa_{\phi}\, .
\end{split}
\ee

\acknowledgments

This work was supported  by  the Natural Sciences and Engineering
Research Council of Canada. The  author is also grateful to the
Killam Trust for its financial support.

\vspace{1.5cm}


%

\end{document}